\documentclass{phbauth}        

\begin{document}

\begin{frontmatter}

\title{
Indirect RKKY Interaction in the Nearly Nested Fermi Liquid
}

\author[address1]{D.N.\ Aristov \thanksref{thank1}},
\author[address1]{S.V.\ Maleyev}

\address[address1]{Petersburg Nuclear Physics Institute, Gatchina 188350
St.Petersburg, Russia }

\thanks[thank1]{Corresponding author. E-mail: aristov@thd.pnpi.spb.ru}

\begin{abstract}
The indirect RKKY interaction in a layered metal with nearly nested
(almost squared) Fermi surface is evaluated analytically. The final
expressions are obtained in closed form as a combination of Bessel
functions.  We show that the expected ``$2k_F$'' oscillations
occur as the far asymptote of our expressions, where a value of
the effective Fermi momentum $k_F$ depends on the direction in
$r-$space.  We demonstrate the existence of the intermediate
asymptote of the interaction which is of the sign-reversal
antiferromagnetic type. This part of RKKY interaction is the
only term surviving in the limit of exact nesting.  A good
accordance of our analytical formulas with numerical findings is
found until the interatomic distances.
\end{abstract}


\end{frontmatter}


\def\bk{{\bf k}}

The Ruderman-Kittel-Kasuya-Yosida interaction (RKKY) \cite{rkky} plays
an important role in the discussion of the interaction of
localized moments in a metal through polarization of conduction
electrons.
It is well known that in three spatial dimension and for the spherical
Fermi surface (FS) this interaction decays as $r^{-3}$ and has the
$2k_F$ oscillations with the Fermi momentum $k_F$.
In the case of anisotropic FSs the RKKY at largest distances
is characterized by the direction-dependent
period of oscillations $(2k_F^\ast)^{-1}$ and can be represented as a
series in $1/r$. \cite{rkky-ani}
Unfortunately, this series diverges at smaller distances
$k_F^\ast r \sim 1$, where the RKKY is mostly significant.

Nowadays the numerical calculations are a principal tool for the
analysis of RKKY in metals with non-spherical FS. However the
theoretical understanding is important here even on the qualitative
level.

In \cite{arimal} we developed an analytical approach fo the calculation
of the RKKY interaction in metals with highly anisotropic FS.

This study
was motivated by the experimental observations, that the rare-earth (R)
subsystem in the high-$T_c$ compounds RBa$_2$Cu$_3$O$_{7-\delta}$
undergoes a transition to the magnetically ordered
antiferromagnetic phase below $\sim 1\,$K. \cite{Lynn}
It can be theoretically shown that the onset of the $d-$wave
superconductivity does not seriously alter the RKKY interaction at
moderate distances \cite{ArMaYa}. The main anisotropy of RKKY below
$T_c$ at those distances is thus determined by the anisotropy of the
FS in the normal state.

To study the possible character of RKKY in high-$T_c$ cuprates, a model
of the nearly-nested (almost squared) FS in a layered metal was chosen.
We analyzed this case in detail and showed a good accordance of our
analytical findings with the numerical results.  It turned out that the
widely used notion of the ``$2k_F$'' oscillations could be applied to
the far asymptote of the interaction.  At intermediate distances the
interaction has the sign-reversal antiferromagnetic character, and only
this behaviour survives at the exact nesting of the FS.

The essence of our method could be outlined as follows.
The range function of the RKKY interaction can be represented as
a sum over the Matsubara frequencies $ \omega_n = \pi T(2n+1)$ :

        \begin{equation}
        \chi({\bf r}) = - T \sum\nolimits_n G(i\omega_n, {\bf r} )^2
	\label{rkk-inter}
 	\end{equation}
with the electronic Green's function
$G(i\omega_n, \bk ) = [i\omega_n - \varepsilon(\bk)]^{-1} $.
Passing here to $r-$representation, we map the whole vicinity of the
FS by the patches of the size $\kappa$ of order of inverse interatomic
distances, where the dispersion $\varepsilon(\bk)$ can be approximated
by a simpler form.
Such mapping can be done for the particular case of the nearly-nested
FS, when the spectrum has a tight-binding form,
$\varepsilon(\bk) = -t( \cos k_x + \cos k_y) + \mu$ and $|\mu|\ll t$.
One distinguishes here the vicinities of the Van Hove points
$(0,\pm\pi)$, $(\pm\pi,0)$ and the flat parts of the spectrum near
$(\pm\pi/2,\pm\pi/2)$.
It is possible to obtain a closed form for the
{\em partial} Green's functions stemmed from each of the patches.
This form is applicable in $r-$space roughly up to the scale
$\kappa^{-1}$.
The total Green's function is estimated as a sum of the partial ones.

The square of Green's function in (\ref{rkk-inter}) yields two types of
contributions. First is the product of the partial Green's
functions $G_j(i\omega_n, {\bf r})^2$ from the same patch $j$.
These terms define the far asymptote of $\chi({\bf r})$,
established earlier, \cite{rkky-ani} and have the
$2k_F^\ast$ oscillations. The second contribution to $\chi({\bf r})$ is
determined by the interference terms
$G_j(i\omega_n, {\bf r}) G_l(i\omega_n, {\bf r})$ with $j\neq l$.
In addition to the smooth dependence on $\bf r$, these latter terms
possess the overall prefactors $\exp i(\bk_j -\bk_l){\bf r}$,
with the spanning vectors $\bk_j -\bk_l$ connecting the centers
of the patches $\bk_j$ and $\bk_l$.

For the case of nearly-nested FS the effective Fermi momentum
$k_F^\ast$ does not exceed the value $\sqrt{|\mu/t|} \ll 1$ and vanishes
along the diagonals $r_x=\pm r_y$.  It turns out that the
interference terms  $G_{(0,\pm\pi)}(i\omega_n, {\bf r})
G_{(\pm\pi,0)}(i\omega_n, {\bf r})$ dominate at the intermediate
distances $r < (k_F^\ast)^{-1}$ for most of the directions of $\bf r$.
The spanning vector between the Van Hove points is ${\bf Q}=
(\pi,\pi)$, thus the commensurate antiferromagnetic modulation
$\exp i {\bf Qr}$ of the RKKY is observed.
Along the diagonals the RKKY interaction is of the ferromagnetic
sign, and non-oscillatory for moderate $r < |t/\mu|$ ; this, however,
does not determine a particular type of magnetic ordering in the
subsystem of localized moments.

Our results suggest that definite shapes of the FS favor the
commensurate antiferromagnetic ordering of localized spins.
Particularly, it should take place, if  the FS lies close to the
symmetry points of the Brillouin zone, the case relevant to the
high-$T_c$ cuprates. Thus our findings could explain the
antiferromagnetism in the subsystem of rare-earth ions, systematically
observed at low temperatures in these compounds.

The partial financial support from the Russian State Program for
Statistical Physics (grant VIII-2) is acknowledged with thanks.



\begin{thebibliography}{9}

\bibitem{rkky} M.A. Ruderman, C. Kittel,
Phys.Rev. {\bf 96}, 99 (1954);
T. Kasuya, Progr.Theoret.Phys.\ (Kyoto) {\bf 16}, 45 (1956);
K. Yosida, Phys.Rev. {\bf 106}, 893 (1957).

\bibitem{rkky-ani} L.M.\ Roth, H.J.\ Zeiger, T.A.\ Kaplan,
Phys.Rev.\ {\bf 149}, 519 (1966).

\bibitem{arimal}
D.N. Aristov and S.V. Maleyev,
 Phys. Rev. B {\bf 56}, 8841 (1997) ; see also
D.N. Aristov, {\em ibid.} {\bf 55}, 8064 (1997).

\bibitem{Lynn} see, e.g., J.W. Lynn, Physica {\bf B 163}, 69 (1990).

\bibitem{ArMaYa}
D.N. Aristov, S.V.  Maleyev, A.G. Yashenkin,
 Z.Phys. B {\bf 102}, 467 (1997).



\end{thebibliography}
\end{document}